\documentclass[onecolumn,useAMS,usenatbib]{mn2e}
\usepackage{graphics,epsfig}% Include figure files
\usepackage{times}% Include figure files
\usepackage{dcolumn}% Align table columns on decimal point
\newcommand{\al}{\ensuremath{\alpha}}
\newcommand{\dal}{\ensuremath{\Delta \alpha/ \alpha}}

\newcommand{\beq}{\begin{equation}}
\newcommand{\eeq}{\end{equation}}
\newcommand{\noi}{\noindent}
\newcommand{\lb}{\left(}
\newcommand{\rb}{\right)}
\newcommand{\lsb}{\left[}
\newcommand{\rsb}{\right]}

\title[The use of OH ``main'' lines to $\ldots$ ]{The use of OH ``main'' lines to constrain the variation of fundamental constants}
\author[Kanekar \& Chengalur]{Nissim Kanekar$^1$\thanks{E-mail: nissim@astro.rug.nl (NK);
chengalu@ncra.tifr.res.in (JNC);}, J. N. Chengalur$^2$\footnotemark[1] \\
$^{1}$ Kapteyn Institute, University of Groningen, Post Bag 800, 9700 AV Groningen, The Netherlands\\
$^{2}$ National Centre for Radio Astrophysics, Post Bag 3, Ganeshkhind, Pune 411 007, India }

\begin{document}

\date{Received mmddyy/ accepted mmddyy}

\maketitle

\label{firstpage}

\begin{abstract}
We describe a new technique to measure variations in the fundamental 
parameters $\alpha$ and $y \equiv m_e/m_p$, using the sum of the frequencies 
of cm-wave OH ``main'' lines. The technique is $\sim$ three orders of 
magnitude more sensitive than that of \cite{chengalur03}, which 
utilised only the four 18cm OH lines. The increase in sensitivity stems from the use 
of OH ``main'' lines 
arising from different rotational states, instead of the frequency difference 
between lines from the same state. We also show that redshifts of the main 
OH 18cm and 6cm lines can be combined with the redshift of an HCO$^+$  transition
to measure any evolution in $\alpha$ and $y$. Both 18cm main lines and a number 
of HCO$^+$ lines have already been detected in absorption in four cosmologically distant
systems; the detection of the main 6cm OH line in any of these systems would thus 
be sufficient to simultaneously constrain changes in $\alpha$ and $y$ between 
the absorption redshift and today.

\end{abstract}

\maketitle

\section{\label{sec:intro} Introduction}

In recent times, quasar absorption lines have emerged as an excellent probe of 
changes in the values of the fundamental ``constants'' 
(e.g.~\citealt{webb99,carilli00,ivanchik03}). Such variations are expected in theories 
like extra-dimensional Kaluza-Klein models or super-string theories, where 
values of the coupling parameters such as the fine structure constant $\alpha$ or 
the gravitational constant $G$ depend on the expectation values of some cosmological 
scalar field(s); changes in these parameters are thus to be expected if the latter 
varies with space and/or time. Various experimental and observational bounds are 
available on the temporal evolution of different coupling constants: these include 
the fine structure constant, $\al$  \citep{ivanchik99,webb01}, the 
gravitational constant $G$ \citep{teller48,damour91}, the combination $g_p \al^2$ 
(where $g_p$ is the proton g-factor; \citealt{drinkwater98,carilli00}), the ratio 
of electron mass to proton mass $y \equiv m_e/m_p$ \citep{ivanchik03}, etc. 
\citet{uzan03} provides a review of the available measurements. 

The most interesting of the new astrophysical estimates are the recent work 
of \cite{webb99,webb01} who claim a detection of changes in the 
numerical value of the fine structure constant $\al$ between high redshift,
$z \sim 3.5$, and the present epoch. The authors initially applied a new 
`many-multiplet' method to absorbers with $1 \la z \la 1.6$ to estimate 
$\dal = (-1.88 \pm 0.53) \times 10^{-5}$ between redshifts $z \sim 1.6$ and 
today \citep{webb99}. This was followed by the use of this method to 
estimate $\dal = (-0.72 \pm 0.18) \times 10^{-5}$ over the redshift 
range $0.5 < z < 3.5$ \citep{webb01} (see, however, \citealt{bekenstein03})
On the other hand, \cite{ivanchik03} constrain the variation in $m_e/m_p$ 
to be $(3.0 \pm 2.4) \times 10^{-5}$ over a similar redshift range  ($ 0< z <3$), 
comparable to the change claimed in the fine structure constant (albeit using a 
different absorber sample). This is somewhat 
surprising, given that most of the above theoretical analyses expect changes 
in different fundamental constants to be coupled: for example, \cite{calmet02} 
and \cite{langacker02} find that variations in the value of $\al$ should be 
accompanied by much larger changes 
(by $\sim 2$ orders of magnitude) in the value of $m_e/m_p$. 

We have recently (\citealt{chengalur03}; hereafter Paper I) demonstrated a new technique to measure 
(or constrain) changes in the fundamental constants using 18cm OH lines. This 
method uses the fact that the four OH lines arise from two very different physical 
phenomena, $\Lambda$-doubling and hyperfine structure, and thus have different 
dependences on the parameters $\alpha$, $y$ and $g_p$. Observations 
of all four OH 18cm transitions in a single cosmologically distant absorber can 
thus be used to simultaneously estimate variations in $y$ and $\alpha$, 
assuming that the proton g-factor remains unchanged (e.g.~\citealt{webb01,carilli00}).
We have also used the linear relationship between OH and HCO$^+$ column densities 
observed both in the Milky Way and in four molecular absorbers out to $z \sim 1$
to argue that HCO$^+$ and OH lines probably arise from the same spatial location 
and are thus unlikely to have velocity offsets relative to each other. The OH 18cm 
redshifts can then be combined with the redshift of a single HCO$^+$ line to 
simultaneously estimate changes in all three fundamental parameters $\alpha$, $g_p$ 
and $y$, {\it in the same object}.

A problem with the above approach is that two of the equations used in the 
analysis (equations~(10) and (11) in Paper~I) involve the separation 
of two line frequencies. The four 18cm lines have rest frequencies of 1665.4018~MHz 
and 1667.3590~MHz (``main'' lines, with $\Delta F = 0$), and 1612.2310~MHz and 
1720.5299~MHz (``satellite'' lines, with $\Delta F = 1$). The separation between the 
main line frequencies is a factor of $\sim 1600$ smaller than the sum of these frequencies 
while the separation between the satellite frequencies is around 30 times smaller 
than the above sum. This implies that the error on the redshift of the two 
frequency differences is worse than the error on the redshift of the sum of 
main line frequencies by the same factors and this large error propagates 
into the estimates of changes in the values of $\alpha$, $y$ and $g_p$. To 
emphasise this point, we note that the error on the sum of the main 
line redshifts in B0218+357 is $\sim 5.6 \times 10^{-6}$, while that on the difference 
between these redshifts is $\sim 6.7 \times 10^{-3}$ (Paper I); it is the latter error 
that dominates the accuracy of the technique in constraining any changes in 
the fundamental parameters (and which hence resulted in the large errors 
in the analysis of the OH, HCO$^+$ and HI lines from the $z \sim 0.6846$ absorber 
towards B0218+357 in Paper~I).

We describe in this Letter a new approach to simultaneously measure changes 
in $\alpha$ and $y$ using OH ``main'' lines arising from different OH rotation
states. This is based on the fact that the exact $\Lambda$-doubling 
frequency split depends on the specific quantum numbers of the state in question 
and has a different dependence on $\alpha$ and $y$ in each state. Since the method 
only uses the sum of different ``main'' line frequencies, it is far
more sensitive than that discussed above (Paper~I), which also uses the difference 
between pairs of measured frequencies. Further, 
the sum of OH main line frequencies does not depend on the proton g-factor 
(as the hyperfine effects cancel out); main lines from any three OH rotation 
states can thus be used to simultaneously measure changes in $y$ and $\alpha$. 
We also show that the main lines from two OH rotation states can be used in 
conjunction with an HCO$^+$ transition to simultaneously measure variations in 
$\alpha$ and $y$. The use of OH main lines to measure changes in the fundamental 
constants has also been discussed by \cite{darling03}; this analysis was, however,
based on a simpler approximation to the OH energy levels and also only considered
variations in the fine structure constant $\alpha$.

\section{The sum of OH ``main'' line frequencies}

The sum of the ``main'' line frequencies $\nu_s$ in an OH rotation state 
essentially gives the energy split due to $\Lambda$-doubling; this can be 
written as 

\beq
\label{eqn:sum1}
\nu_s\;\; = \;\; q_\Lambda\lb J+1/2 \rb\lsb 
\lb 2 + \frac{A'}{B'} \rb \lb 1 + \frac{2 - A/B}{X} \rb +
\frac{4 \lb J + 3/2 \rb \lb J - 1/2 \rb}{X} \rsb \;\; ,
\eeq

\noi where $J$ is the rotational quantum number, $A$, the fine structure interaction 
constant and $B$, the rotation constant \citep{vanvleck29,townes55}. The quantity $X$ is 
defined by 
\beq
\label{eqn:x}
X = \pm \lsb \lb A/B \rb \{ {\lb A/B\rb} - 4 \}+ 4\lb J + 1/2 \rb^2 \rsb^{1/2} \;\; ,
\eeq
\noi with the negative sign for the $^2\Pi_{3/2}$ state and the positive sign for 
the $^2\Pi_{1/2}$ state \citep{townes55}. Further, $q_\Lambda \approx 4B^2/h\nu_e$,
where $h\nu_e$ is the energy difference between the ground and first excited 
electronic state. Numerically, $A/B = -7.547$ and $A'/B' = -6.073$ \citep{townes55}. 
The above quantities have the following dependences on the fundamental constants 
$\alpha$, $y$ and  $R_\infty$ : $A' \propto A \propto \alpha^2 R_\infty$, 
$B' \propto B \propto y R_\infty$, where $R_\infty$ is the Rydberg constant.
For the rotation constant $B$,
we have assumed, as usual (e.g.~\citealt{murphy01}), that variations in $(m_p/M)$,
which are suppressed by a factor $m_p/U \sim 100$ (where $M$ is the reduced mass and
U the binding energy) can be ignored. We thus have $ \lsb A'/B' \rsb \propto \lsb A/B 
\rsb \propto \lb \alpha^2/y \rb $. Finally, we note that equation~(\ref{eqn:sum1})
does not depend on the proton g-factor $g_p$.

Replacing the above scalings in equation~(\ref{eqn:sum1}) for $\nu_s$, we obtain
$\nu_s \propto y^2 R_\infty F\lb \alpha^2/y\rb \;\;$, where $F \equiv F \lb \beta \rb $
is a function which depends only on the ratio $\beta \equiv A/B \propto \alpha^2/y$ and 
is defined by
\beq
F\lb \beta \rb = \lsb \lb 2 + \frac{6.073}{7.547}\beta \rb \lb 1 +
\frac{2 - \beta}{X\lb\beta\rb} \rb + \frac{4 \lb J + 3/2 \rb \lb J - 1/2 \rb}{X\lb \beta\rb} \rsb \; \; .
\eeq

\noi Thus, the ratio of the change in the sum of any two main line frequencies 
$\Delta \nu_s$ to their sum today is given by
\begin{eqnarray}
\frac{\Delta \nu_s }{\nu_s} &=& 2\frac{\Delta y}{y} + \frac{\Delta R_\infty}{R_\infty}
+ \frac{\Delta F \lb \beta \rb} {F\lb\beta \rb}\\
\label{eqn:sum}
&=& 2\frac{\Delta y}{y} + \frac{\Delta R_\infty}{R_\infty} 
+ \frac{\beta}{F} \frac{dF}{d\beta} \lsb 2 \frac{\Delta \alpha}{\alpha} - 
\frac{\Delta y}{y} \rsb 
\end{eqnarray}

\beq
\mathrm{where} \;\;\;\; \frac{dF}{d\beta} = C \lb 1 + \frac{2 - \beta}{X} \rb 
- \frac{2 + C\beta}{X}\lsb 1 + \frac{2 - \beta}{X} \frac{dX}{d\beta} \rsb 
- \frac{4 \lb J + 3/2 \rb \lb J - 1/2 \rb}{X^2} \frac{dX}{d\beta} \;\;.
\eeq
\noi In the above, $dX/d\beta = \lb \beta - 2\rb/X$ and $C = (6.073/7.547)$. 
\noi Equation~(\ref{eqn:sum}) has been evaluated for the main lines of the 
$^2\Pi_{3/2}, J = 3/2$ state in Paper~I; we 
extend this analysis to the main lines of other rotational states that have been 
detected in the Galactic interstellar medium. The results are listed below, 
along with the rest frequencies of the ``main'' lines for each state :

\begin{enumerate}
\item{$\mathbf{^2\Pi_{3/2}, J = 3/2}$ : Rest frequencies : 1667.3590 and 1665.4018~MHz 
(i.e. $\sim $ 18cm). Equation~(\ref{eqn:sum}) yields 
\beq
\label{eqn:18cm}
\frac{\Delta \nu_s }{\nu_s} = 2.571 \frac{\Delta y}{y} -
1.141 \frac{\Delta \alpha}{\alpha} + \frac{\Delta R_\infty}{R_\infty}
\eeq
}
\item{$\mathbf{^2\Pi_{1/2}, J = 1/2}$ : This state has a single main line, at a rest 
frequency of 4750.656~MHz ($\sim $ 6cm); the second main line corresponds to
an F = 0--0 transition, which is forbidden by the selection rules. This might appear 
to be a drawback, since it would seem impossible to detect both main lines. However, 
it turns out that the frequency separation between the two main lines vanishes, to first 
order. This can be seen as the separation between the two main line frequencies 
$\nu_{1}$ and $\nu_{2}$ is 
\beq
\label{eqn:diff1}
\Delta \nu\;\; \equiv\;\; \nu_{1} - \nu_{2}\;\; =\;\; \frac{-2d \lb -X - 2 + \beta \rb }{3 X}  
\eeq
\noi where $d$ is a hyperfine constant (equation~(46) of \citealt{dousmanis55}). Further, 
$X = \beta -2$ for $J = 1/2$, implying that $\Delta \nu = 0$. Thus, the sum of the two 
main line frequencies for the $J = 1/2$ case is, to first order, merely equal to twice
the F = 1--1 frequency and a detection of this transition is sufficient to measure 
the above sum. In this case, equation~(\ref{eqn:sum}) yields 
\beq
\label{eqn:6cm}
\frac{\Delta \nu_s }{\nu_s} = 0.509 \frac{\Delta y}{y} +
2.982 \frac{\Delta \alpha}{\alpha} + \frac{\Delta R_\infty}{R_\infty}
\eeq
}
\item{$\mathbf{^2\Pi_{3/2}, J = 5/2}$ : Rest frequencies : 6030.747~MHz and 6035.092~MHz 
($\sim $ 5cm). Equation~(\ref{eqn:sum}) yields 
\beq
\label{eqn:5cm}
\frac{\Delta \nu_s }{\nu_s} = 2.452 \frac{\Delta y}{y} -
0.903 \frac{\Delta \alpha}{\alpha} + \frac{\Delta R_\infty}{R_\infty}
\eeq
}
\item{$\mathbf{^2\Pi_{1/2}, J = 3/2}$ : Rest frequencies : 7761.747~MHz and 7820.125~MHz 
($\sim $ 3.8cm). Equation~(\ref{eqn:sum}) yields 
\beq
\label{eqn:3.8cm}
\frac{\Delta \nu_s }{\nu_s} = 0.072 \frac{\Delta y}{y} +
3.857 \frac{\Delta \alpha}{\alpha} + \frac{\Delta R_\infty}{R_\infty}
\eeq
}
\item{$\mathbf{^2\Pi_{1/2}, J = 5/2}$ : Rest frequencies : 8135.870~MHz and 8159.587~MHz 
($\sim $ 3.7cm). Equation~(\ref{eqn:sum}) yields 
\beq
\label{eqn:3.7cm}
\frac{\Delta \nu_s }{\nu_s} = -0.920 \frac{\Delta y}{y} +
5.840 \frac{\Delta \alpha}{\alpha} + \frac{\Delta R_\infty}{R_\infty}
\eeq
}
\item{$\mathbf{^2\Pi_{3/2}, J = 7/2}$ : Rest frequencies : 13434.596~MHz and 
13441.4173~MHz ($\sim $ 2.2cm). Equation~(\ref{eqn:sum}) yields 
\beq
\label{eqn:2.2cm}
\frac{\Delta \nu_s }{\nu_s} = 2.334\frac{\Delta y}{y} -
0.678 \frac{\Delta \alpha}{\alpha} + \frac{\Delta R_\infty}{R_\infty}
\eeq
}
\end{enumerate}

We note that each of the above equations~(\ref{eqn:18cm} -- \ref{eqn:2.2cm}) has 
the same dependence on the Rydberg constant $R_\infty$, but different dependences on 
$y$ and $\alpha$. If we have two transitions whose rest frequencies $\nu_1(0)$ and 
$\nu_2(0)$ depend on redshift, due to the evolution of various fundamental 
constants such as $\al$, $y$, etc, the first order difference between the measured 
redshifts is (e.g.~Paper~I)
\beq 
\label{eqn:redshift}
\frac {\Delta z }{ 1 + {\bar z} }\;\; =\;\; \lsb \frac{ \Delta \nu_2}{\nu_2(0)} \rsb - 
\lsb \frac{\Delta \nu_1}{\nu_1(0)} \rsb \;\; ,
\eeq

\noi where $\bar z$ is the mean measured redshift. Given two spectral lines (or 
linear combinations of line frequencies) with different dependences on some 
fundamental parameter, the differences between the measured redshifts can be used 
to constrain the evolution of the parameter in question. Clearly, any three 
of the equations~(\ref{eqn:18cm} -- \ref{eqn:2.2cm}) can be combined in pairs
in equation~(\ref{eqn:redshift}) to yield two simultaneous equations 
in $\Delta \alpha / \alpha$ and $\Delta y / y$, which can then be solved to 
measure any changes in both these quantities. Thus, the detection of the ``main''
OH lines in any three of the OH rotation states in a single absorber can be used 
to simultaneously measure changes in $\alpha$ and $y \equiv m_e/m_p$ at the same
physical space-time location. Note that the large number of OH rotational 
states allows a simple self-consistency check of any such measurement, by a search 
for the ``main'' lines from a fourth OH rotation state. Since this technique 
uses different lines from the same species, systematic velocity offsets between
the different absorption lines are unlikely to be the dominant source of error.
It should be noted, however, that the higher OH rotational levels may be 
excited in regions with very different physical conditions from regions 
giving rise to OH ground state absorption; the possibility of velocity
offsets between the 18cm absorbing gas and the 6cm or 5cm absorbing gas 
hence cannot be ruled out. However, the OH column density of the absorbing 
gas can be independently estimated from the different OH lines; if these estimates 
are found to be in agreement, it would argue that all the lines originated in the same gas cloud.

HCO$^+$ rotational transitions also have no dependence on the proton 
g-factor $g_p$, with line frequencies proportional to $y R_\infty$. Moreover, 
as discussed in Paper~I, the linear relation between HCO$^+$ and OH 
column densities observed both in the galaxy \citep{liszt96} and out to $z \sim 1$ 
\citep{kanekar02} (extending over more than two orders of magnitude in column 
density) suggests that the two species are likely to be located in the same 
region of a molecular cloud. This implies that one of the OH rotation states 
in the above analysis can be replaced with an HCO$^+$ transition. Both 18cm 
main lines and a number of HCO$^+$ transitions have already been detected in 
four molecular absorbers between $z \sim 0.25$ and $z \sim 0.9$ 
\citep{wiklind95,wiklind96a,wiklind96b,wiklind97,chengalur99,kanekar02,kanekar03}; the 
detection of a single 6cm ``main'' line in one of these absorbers (which seems 
the most promising of the remaining OH main lines) would thus be sufficient to carry 
out the above measurement. We end this discussion by setting out the two simultaneous 
equations obtained by using the 18cm and 6cm main lines and an HCO$^+$ line
in equation~(\ref{eqn:redshift}). Combining the sum of 18cm main line frequencies 
with the HCO$^+$ frequency gives
\beq
\label{eqn:18cm-HCO}
\frac {\Delta z_{13}}{ 1 + {\bar z_{13}} }\;\; =\;\; 1.571 \frac{\Delta y}{y} -
1.141 \frac{\Delta \alpha}{\alpha} \;\; .
\eeq

\noi Similarly, combining the frequencies of the OH 6cm main line with an HCO$^+$ 
transition yields
\beq
\label{eqn:6cm-HCO}
\frac {\Delta z_{23}}{ 1 + {\bar z_{23}} }\;\; =\;\; -0.491\frac{\Delta y}{y} +
2.983 \frac{\Delta \alpha}{\alpha} \;\; ,
\eeq

\noi where $\Delta z_{ij} = z_{j} - z_{i}$, ${\bar z}_{ij} = \lb z_i + z_j \rb/2$ 
and the subscripts 1, 2 and 3 denote 
the 18cm, 6cm and HCO$^+$ transitions respectively. A detection of the main 6cm OH 
lines in any of the four cosmologically distant absorbers can thus be immediately 
used in the above equations to measure $\dal$ and $\Delta y/y$. 

It has been emphasized in the introduction that the primary drawback of our 
earlier analysis, using the 18cm OH lines (Paper~I), is that the frequency 
separation between the main 18cm lines is only $\sim 1.957/(1 + z )$~MHz, where 
$z$ is the absorption redshift. This dominated the errors in redshift measurements 
and hence resulted in the large final errors in the analysis of the absorption lines 
towards B0218+357. Of course, the advantage of this method is that it simultaneously 
allows a measurement of changes in three fundamental parameters, $g_p$, $\alpha$ and 
$y \equiv m_e/m_p$. In the present case, while we have a far higher accuracy in the 
measurement (as the frequency sum of the 18cm main lines is larger than their frequency 
difference by a factor of $\sim 1600$ and by an even higher factor for the 
higher order transitions), the technique only allows an estimate of changes in $\alpha$ 
and $y$. However, these estimates can be replaced in the equation for the separation 
between the satellite 18cm OH lines (equation~(11) in Paper~I) to constrain the 
evolution of the proton g-factor $g_p$. While this last measurement would be a 
factor of $\sim 30$ less sensitive than the measurements of $\dal$ and $\Delta y/ y$, 
it is still  more than 50 times more sensitive than estimates of $\Delta g_p/ g_p$ 
obtained using the earlier method (Paper~I); this is due to the fact that the 
frequency separation between the 18cm satellite lines is $\sim 108/(1+z)$~MHz, fifty times 
larger than that between the 18cm main lines.

We note, in passing, that other ``Lambda-doubled'' systems are known to exist in the 
laboratory and any of these could, in principle, be used in place of OH in a similar 
calculation.  However, to the best of our knowledge, multiple transitions have not been 
detected in these other systems in astrophysical sources; we suspect that the strength 
of cm-wave OH lines is likely to make OH the best candidate for such analyses. Further, 
the present calculation is based on a perturbative treatment of the OH levels 
\citep{vanvleck29, townes55}; more recent analyses (e.g.~\citealt{brown79}) use the 
``effective Hamiltonian'' approach, resulting in higher order effects. As pointed 
out in Paper~I, these are unlikely to significantly affect our results.

Finally, we estimate the accuracy that could be obtained in a measurement of 
$\Delta \alpha / \alpha$ by the present method, for the $z \sim 0.885$ absorber 
towards PKS~1830$-$21; the latter has the highest redshift of all known 
18cm OH absorbers \citep{kanekar02}. We simplify the analysis by assuming that $y$ is constant, so that 
main lines of only two rotational states need be used; we will consider the 
$\mathbf{^2\Pi_{3/2}, J = 3/2}$ and $\mathbf{^2\Pi_{1/2}, J = 1/2}$ states, i.e. 
main lines at observing frequencies of $\sim 885$~MHz and $\sim 2520$~MHz, 
respectively, for an absorber at $z = 0.885$. Next, a resolution of 2~kHz is 
not unreasonable for radio spectroscopy with present-generation telescopes. Even 
if we assume that the line centroids are only determined to this accuracy (i.e. that 
sub-channel resolution is not obtained, via fitting to the line profile), the errors 
on the redshifts of the sum of main line frequencies would be $\Delta z = 1 \times 10^{-6}$ 
and $\Delta z = 4 \times 10^{-7}$, for the 18cm and 6cm lines, respectively. Combining 
equations~(\ref{eqn:18cm}) and (\ref{eqn:6cm}) in equation~(\ref{eqn:redshift}) (and assuming 
$y$ to be constant) then gives $\Delta \alpha / \alpha = 1.5 \times 10^{-7}$,
the $1 \sigma$ accuracy in a measurement of changes in $\alpha$ from $z \sim 0.885$ 
to today. This is significantly better than the precision obtained in the best optical 
studies today (e.g. $\Delta \alpha / \alpha = 6 \times 10^{-7}$; \citet{humchand04}). 
Note, however, that the above errors do not take into account systematic effects, i.e. 
the possibility of relative motions between the 18cm and 6cm absorbing clouds; as 
mentioned earlier, observations of lines from other OH rotational states would help 
to constrain such systematics.

In summary, we have demonstrated a new technique to simultaneously measure any
evolution in two fundamental constants $\alpha$ and $y \equiv m_e/m_p$, using OH 
``main'' absorption lines. The method is $\sim$ three orders of magnitude more 
sensitive than that described by Chengalur \& Kanekar 
(2003), which utilised the four OH 18cm lines for the analysis. The increase in 
sensitivity comes from the use of OH ``main'' lines arising from different OH 
rotational states, rather than the difference between the frequencies of lines 
arising from the same state. The technique requires the detection of main lines 
from three OH rotational states or, alternately, two OH states and one HCO$^+$ 
transition, in the same absorber. The large number of OH rotational states also 
allows a simple consistency check of any measurement by a search for the main 
lines of one further OH state. Both 18cm main lines and a number of HCO$^+$ 
transitions have been detected in absorption in four extra-galactic molecular 
absorbers. The detection of a single 6cm main line in any of these systems would 
thus be sufficient to simultaneously measure (or constrain) changes in $\alpha$ 
and $y \equiv m_e/m_p$.

\section{Acknowledgments}
	We are grateful to Rajaram Nityananda for very useful discussions
on the energy levels of the OH ground state.

\bibliographystyle{mn2e}
\bibliography{ms}

\begin{thebibliography}{}

\bibitem[\protect\citeauthoryear{Bekenstein}{Bekenstein}{2003}]{bekenstein03}
Bekenstein J.~D.,  2003, astro-ph/0301566

\bibitem[\protect\citeauthoryear{Brown \& Merer}{Brown \&
  Merer}{1979}]{brown79}
Brown J.~M.,  Merer A.~J.,  1979, J. Mol. Spectr, 74, 488

\bibitem[\protect\citeauthoryear{Calmet \& Fritzsch}{Calmet \&
  Fritzsch}{2002}]{calmet02}
Calmet X.,  Fritzsch H.,  2002, Phys. Lett. B, 540, 173

\bibitem[\protect\citeauthoryear{Carilli, Menten, Stocke, Perlman, Vermeulen,
  Briggs, de Bruyn, Conway \& Moore}{Carilli et~al.}{2000}]{carilli00}
Carilli C.~L.,  Menten K.~M.,  Stocke J.~T.,  Perlman E.,  Vermeulen R.,
  Briggs F.,  de Bruyn A.~G.,  Conway A.,    Moore C.~P.,  2000, Phys. Rev.
  Lett., 85, 5511

\bibitem[\protect\citeauthoryear{Chand, Srianand, Petitjean \& Aracil}{Chand
  et~al.}{2004}]{humchand04}
Chand H.,  Srianand R.,  Petitjean P.,    Aracil B.,  2004, A\&A, 417, 853

\bibitem[\protect\citeauthoryear{Chengalur, de Bruyn \& Narasimha}{Chengalur
  et~al.}{1999}]{chengalur99}
Chengalur J.~N.,  de Bruyn A.~G.,    Narasimha D.,  1999, A\&A, 343, L79

\bibitem[\protect\citeauthoryear{Chengalur \& Kanekar}{Chengalur \&
  Kanekar}{2003}]{chengalur03}
Chengalur J.~N.,  Kanekar N.,  2003, Phys. Rev. Lett., 91, 241302

\bibitem[\protect\citeauthoryear{Damour \& Taylor}{Damour \&
  Taylor}{1991}]{damour91}
Damour T.,  Taylor J.~H.,  1991, Astrophys. J., 366, 501

\bibitem[\protect\citeauthoryear{Darling}{Darling}{2003}]{darling03}
Darling J.,  2003, Phys. Rev. Lett., 91, 011301

\bibitem[\protect\citeauthoryear{Dousmanis, Sanders \& Townes}{Dousmanis
  et~al.}{1955}]{dousmanis55}
Dousmanis G.~C.,  Sanders T.~M.,    Townes C.~H.,  1955, Phys. Rev., 100, 1735

\bibitem[\protect\citeauthoryear{Drinkwater, Webb, Barrow \&
  Flambaum}{Drinkwater et~al.}{1998}]{drinkwater98}
Drinkwater M.~J.,  Webb J.~K.,  Barrow J.~D.,    Flambaum V.~V.,  1998, Mon.
  Not. R. Astron. Soc., 295, 457

\bibitem[\protect\citeauthoryear{Ivanchik, Potekhin \& Varshalovich}{Ivanchik
  et~al.}{1999}]{ivanchik99}
Ivanchik A.~B.,  Potekhin A.~Y.,    Varshalovich D.~A.,  1999, A\&A, 343, 439

\bibitem[\protect\citeauthoryear{Ivanchik, Petitjean, Rodriguez \&
  Varshalovich}{Ivanchik et~al.}{2003}]{ivanchik03}
Ivanchik A.~V.,  Petitjean P.,  Rodriguez E.,    Varshalovich D.~A.,  2003,
  Ap\&SS, 283, 583

\bibitem[\protect\citeauthoryear{Kanekar \& Chengalur}{Kanekar \&
  Chengalur}{2002}]{kanekar02}
Kanekar N.,  Chengalur J.~N.,  2002, A\&A, 381, L73

\bibitem[\protect\citeauthoryear{Kanekar, Chengalur, de Bruyn \&
  Narasimha}{Kanekar et~al.}{2003}]{kanekar03}
Kanekar N.,  Chengalur J.~N.,  de Bruyn A.~G.,    Narasimha D.,  2003, MNRAS
  (Letters), 345, L7

\bibitem[\protect\citeauthoryear{Langacker, Segr\'{e} \& Strassler}{Langacker
  et~al.}{2002}]{langacker02}
Langacker P.~G.,  Segr\'{e} G.,    Strassler M.~J.,  2002, Phys. Lett. B, 528,
  121

\bibitem[\protect\citeauthoryear{Liszt \& Lucas}{Liszt \&
  Lucas}{1996}]{liszt96}
Liszt H.,  Lucas R.,  1996, A\&A, 314, 917

\bibitem[\protect\citeauthoryear{Murphy, Webb, Flambaum, Drinkwater, Combes \&
  Wiklind}{Murphy et~al.}{2001}]{murphy01}
Murphy M.~T.,  Webb J.~K.,  Flambaum V.~V.,  Drinkwater M.~J.,  Combes F.,
  Wiklind T.,  2001, MNRAS, 327, 1244

\bibitem[\protect\citeauthoryear{Teller}{Teller}{1948}]{teller48}
Teller E.,  1948, Phys. Rev., 73, 801

\bibitem[\protect\citeauthoryear{Townes \& Schawlow}{Townes \&
  Schawlow}{1955}]{townes55}
Townes C.~H.,  Schawlow A.~L.,  1955, Microwave Spectroscopy, McGraw-Hill Book
  Company, U.S.A.

\bibitem[\protect\citeauthoryear{Uzan}{Uzan}{2003}]{uzan03}
Uzan J.-P.,  2003, Rev. Mod. Phys, 75, 403

\bibitem[\protect\citeauthoryear{{Van Vleck}}{{Van Vleck}}{1929}]{vanvleck29}
{Van Vleck} J.~H.,  1929, Phys. Rev., 33, 467

\bibitem[\protect\citeauthoryear{Webb, Flambaum, Churchill, Drinkwater \&
  Barrow}{Webb et~al.}{1999}]{webb99}
Webb J.~K.,  Flambaum V.~V.,  Churchill C.~W.,  Drinkwater M.~J.,    Barrow
  J.~D.,  1999, Phys. Rev. Lett, 82, 884

\bibitem[\protect\citeauthoryear{Webb, Murphy, Flambaum, Dzuba, Barrow,
  Churchill, Prochaska \& Wolfe}{Webb et~al.}{2001}]{webb01}
Webb J.~K.,  Murphy M.~T.,  Flambaum V.~V.,  Dzuba V.~A.,  Barrow J.~D.,
  Churchill C.~W.,  Prochaska J.~X.,    Wolfe A.~M.,  2001, Phys. Rev. Lett,
  87, 091301

\bibitem[\protect\citeauthoryear{Wiklind \& Combes}{Wiklind \&
  Combes}{1995}]{wiklind95}
Wiklind T.,  Combes F.,  1995, A\&A, 299, 382

\bibitem[\protect\citeauthoryear{Wiklind \& Combes}{Wiklind \&
  Combes}{1996a}]{wiklind96a}
Wiklind T.,  Combes F.,  1996a, A\&A, 315, 86

\bibitem[\protect\citeauthoryear{Wiklind \& Combes}{Wiklind \&
  Combes}{1996b}]{wiklind96b}
Wiklind T.,  Combes F.,  1996b, Nature, 379, 139

\bibitem[\protect\citeauthoryear{Wiklind \& Combes}{Wiklind \&
  Combes}{1997}]{wiklind97}
Wiklind T.,  Combes F.,  1997, A\&A, 328, 48

\end{thebibliography}

\end{document}